\documentclass[aps,nofootinbib,prl,superscriptaddress,tightenlines,notitlepage,
twocolumn,showpacs,floatfix]{revtex4-1}
\usepackage[colorlinks]{hyperref}
\usepackage{tabularx}
\usepackage{bm}
\usepackage{graphicx}
\usepackage{amssymb,bm,tensor}
\usepackage{xcolor}
\usepackage{amsmath}
\usepackage[varg]{txfonts}
\usepackage{enumerate}
\usepackage[normalem]{ulem}
\usepackage{bm}
\usepackage[OT1]{fontenc}
\usepackage{amsmath}
\usepackage{makecell} 

\newcommand{\alosgal}{\ensuremath{a_{\rm los,gal}}}
\newcommand{\alosgalmod}{\ensuremath{a_{\rm los,gal}^{\rm mod}}}
\newcommand{\alosshmod}{\ensuremath{a_{\rm los,sub-halo}^{\rm mod}}}

\newcommand{\PBDOTobs}{\ensuremath{\dot{P}_b^{\rm Obs}}}
\newcommand{\PBDOTshk}{\ensuremath{\dot{P}_b^{\rm Shk}}}
\newcommand{\PBDOTgal}{\ensuremath{\dot{P}_b^{\rm Gal}}}
\newcommand{\PBDOTGR}{\ensuremath{\dot{P}_b^{\rm GR}}}

\newcommand{\araa}{ARAA}
\newcommand{\jcap}{JCAP}
\newcommand{\mnras}{MNRAS}
\newcommand{\apjl}{ApJL}
\newcommand{\aap}{AAP}

\begin{document}

\title{Constraints on a dark matter sub-halo near the Sun from pulsar timing}

\date{\today}

\author{Sukanya Chakrabarti}
\affiliation{Department of Physics \& Astronomy, University of Alabama, Huntsville, Huntsville, AL 35899}\email{sc0236@uah.edu}

\author{Philip Chang}
\affiliation{Department of Physics \& Astronomy, University of Wisconsin-Milwaukee, Milwaukee, Wisconsin 53211}

\author{Stefano Profumo}
\affiliation{Department of Physics, University of Calfornia, Santa Cruz, 95064}

\author{Peter Craig}
\affiliation{Department of Physics \& Astronomy, Michigan State University, East Lansing, MI 48824}

\begin{abstract}
 Using pulsar accelerations, we identify and constrain the properties of a dark matter sub-halo in the Galaxy for the first time from analyzing the acceleration field of binary and solitary pulsars.  The sub-halo is characterized by analyzing a local deviation from a smooth potential.  Our MCMC calculations show that this sub-halo has a mass of 
 $2.45^{+1.07}_{-0.96} \times 10^{7}~M_{\odot}$ and is located at Galactocentric coordinates $X = 7.43^{+0.2}_{-0.12}~\rm$ kpc, $Y = 0.38^{+0.11}_{-0.16} ~\rm kpc$, $Z = 0.21^{+0.06}_{-0.11} ~\rm kpc$, using flat, uninformative priors, where we have modeled the sub-halo as a compact object. The Bayes factors for the models are in the range of $\sim$ 20-40, which indicates tentative evidence (though not yet decisive) for the sub-halo.  Modeling the sub-halo with a NFW profile gives a sub-halo mass of $6.19^{+1.92}_{-2.03} \times 10^{7} M_{\odot}$,  located at $X = 7.47^{+0.21}_{-0.14}$, $Y=0.38^{+0.11}_{-0.16}$,
$Z=0.21^{+0.06}_{-0.11}$; the mass within the scale radius (0.1 kpc) for the NFW profile is $0.48^{0.15}_{-0.16} \times 10^{7} M_{\odot}$.
 We examine \textit{Gaia} data and the atomic and molecular hydrogen data of our Galaxy and show that the measured deviation from a smooth potential cannot arise from the gas or the stars in our Galaxy.
 Additionally, by analyzing the full sample of binary pulsars with available acceleration measurements that span 3.4 kpc in Galactocentric radius from the Sun and 3.6 kpc in vertical height, we find that massive (with mass $>10^{8}~M_{\odot}~$) sub-halos are disfavored for the Milky Way within several kiloparsec of the Sun.  
 While smaller sub-halos may be present, they are beyond the reach of current direct acceleration measurements.  The presence of a $\sim 10^{7}~M_{\odot}$ sub-halo within a few kpc of the Sun is potentially consistent with the expected number counts of sub-halos in the prevailing $\Lambda$CDM paradigm, for a substantial sub-halo mass fraction.  
 As the number and precision of direct acceleration measurements continues to grow, we will obtain tighter constraints on dark matter sub-structure in our Galaxy.  This work now provides a proof of principle for probing nearby, low-mass sub-halos, and has implications across many fields of astrophysics - from understanding the nature of dark matter to galaxy formation.

\end{abstract}

\maketitle

\textit{Introduction.}---
Dark matter sub-structure, i.e., dark sub-halos, are the cornerstone of dark matter models \cite{Bullock_BoylanKolchin2017}.  The prevailing $\Lambda$ CDM paradigm, as established in dark-matter-only simulations such as Via Lactea II and Aquarius, predicts that dark matter subhalos should be abundant in Milky Way–mass halos, following a steep mass function toward low mass. While baryonic effects suppress the survival of low-mass subhalos in the inner Galaxy, simulations that include realistic baryonic physics still find that subhalos with masses $\geq 10^{7} M_{\odot}$ can survive at distances $\ge 1$ kpc from the Galactic center \cite{Sawalaetal2017,GarrisonKimmeletal2017}. However, finding dark matter sub-halos in the Galaxy has proven challenging and as of yet there are no definitive detections.  
Sub-halos in the mass range $\sim 10^{8}-10^{9}~M_{\odot}$ have been identified in some external galaxies that are gravitational lenses \cite{Vegetti2012,Hezaveh2016}.  In the Milky Way, gaps in stellar streams have been analyzed in a statistical sense to characterize dark matter sub-halos \cite{Erkal2017}.  Most recently, with the release of \textit{Gaia} data, it has become possible to identify individual gaps with increasing fidelity \cite{APW_Bonaca2018,Bonaca2019}.  However, the sub-halo is still characterized via models in a qualitative way leading to a wide range of possible masses (covering several orders of magnitude) for the sub-halo \cite{APW_Bonaca2018,Bonaca2019}. 

Traditionally, astronomers have \emph{estimated} the accelerations of stars in the Galaxy by modeling a snapshot of the positions and velocities of stars \cite{BT08}.  However, \emph{measuring} the accelerations of stars that live within the gravitational potential of the Galaxy using extremely precise time-series observations provides the most direct probe of the mass distribution (the stars and the dark matter).
Here, we employ pulsar timing observations to identify and quantitatively characterize the properties of dark matter sub-halos near the Sun.  Earlier work derived fundamental Galactic parameters that describe the Milky Way using pulsar accelerations \cite{Chakrabartietal2021,Donlonetal2024,Moran2024,Donlonetal2025}, including the average mid-plane density, the slope of the rotation curve and Oort constants, the shape of the Galactic potential, and the local dark matter density.  A key advantage of using direct acceleration measurements to characterize the Galaxy is that one does not have to assume equilibrium or symmetry as is done in traditional kinematic analyses.  Perturbations from orbiting dwarf galaxies have rendered our Galaxy out of equilibrium, leading to large perturbations both in the gas \cite{LevineBlitzHeiles2006, Chakrabarti_Blitz2009} and in the stars \cite{Purcell2011,DeLaVega2015,antoja2018,helmi2018,Chakrabarti2019,Hunt_Vasiliev2025}.  
The assumptions of time-independence and symmetry in traditional kinematic analysis are likely unrealistic given the growing evidence for disequilibrium phenomena in the Galaxy \cite{Haines2019,Hunt_Vasiliev2025}.  Moreover, acceleration measurements represent the instantaneous state of the Galaxy, while phase-space measurements are an integral over these accelerations and therefore will manifest a smoother response.

In this {\it Letter}, we develop a novel method for constraining the properties of dark matter sub-halos in the Milky Way from direct acceleration measurements of pulsars.  Earlier work provides line-of-sight (LOS) Galactic acceleration measurements that constrains the galactic potential from binary pulsars \cite{Chakrabartietal2021,Moran2024,Donlonetal2024}, and we review the basic idea here.
For a binary system in the Galaxy (we note our further specifications in the Appendix), we may write the observed orbital period drift rate $\dot{P}_{b}^{\rm Obs}$ as: 
\begin{equation}\label{eq:Pdot}
   \PBDOTobs = \PBDOTgal -  \PBDOTshk - \PBDOTGR,
\end{equation}
where $\PBDOTgal = P_{b} \alosgal/c$ is the period drift rate induced by the Galactic potential, $\alosgal$ is the relative LOS Galactic acceleration between the solar system and the pulsar, $P_b$ is the period of the binary,
$c$ is the speed of light, and $\PBDOTshk$ is the apparent  drift rate caused by the binary's transverse motion (known as the Shklovskii effect) which is given by: 
\begin{equation}
\PBDOTshk= 
\mu^2 d \frac{P_{b}}{c}, \end{equation}
for a system at distance $d$ with a proper motion $\mu$ \cite{DamourTaylor1991}. The $\PBDOTGR$ term expresses the rate at which the binary pulsar is losing energy due to gravitational radiation \cite{Weisberg2016}, and can be calculated given $P_{b}$, the eccentricity $e$, and the masses of the pulsar $m_p$ and its companion $m_c$ as
\begin{eqnarray}
    \PBDOTGR &=& -\frac{192\pi G^{5/3}}{5c^5} \left(\frac{P_b}{2\pi}\right)^{-5/3} (1-e^2)^{-7/2} \nonumber \\
        && \times \left( 1 + \frac{73}{24}e^2 + \frac{37}{96} e^4 \right) \frac{m_pm_c}{(m_p+m_c)^{1/3}} \,.
\end{eqnarray}
An alert reader will recognize that in equation (\ref{eq:Pdot}) that we implicitly assume that the \PBDOTobs  $~$ignores period drifts from gravitational accelerations that do not come from the smooth Galactic potential, e.g., dark matter subhalos.  The central theme of this work is relaxing this assumption.
Given these terms, one can then solve for the LOS Galactic acceleration, $\alosgal$ as: 
\begin{equation}
\alosgal = c\frac{\dot{P}_{b}^{\rm Gal}}{P_{b}}.
\end{equation}
as in prior work \cite{Chakrabartietal2021,Donlonetal2024}
that accounts for uncertainties in measured values of the distance, proper motion, binary period and period drift, as well as the masses and eccentricities of the pulsars.  A similar method has recently been used to measure Galactic accelerations using solitary pulsars \cite{Donlonetal2025} that uses an empirical calibration of the magnetic braking term, and we employ these recent measurements here as well.  Here we include the scatter in the magnetic braking term as an additional error term as noted in the Appendix of \cite{Donlonetal2025}.

\textit{Constraints on dark matter sub-halos}---Following this work which provides the LOS Galactic acceleration from the observed period drift of binary pulsars \cite{Chakrabartietal2021,Donlonetal2024}, we can now determine the deviation in the acceleration field relative to a smooth gravitational potential using a collection of binary pulsars in close proximity to each other, e.g., breaking the implicit assumption discussed above.  For a set of pulsars at 3-d positions $\bf{x_{i}}$, the deviation in the observed line-of-sight acceleration experienced by each binary pulsar from a smooth potential is: 
\begin{equation}
\Delta a(\mathbf{r}_i) = a_{\rm los,measured}(\mathbf{r}_i) - \alosgalmod(\mathbf{r}_i), 
\end{equation}
where $\rm a_{\rm los, measured}(\mathbf{r})$ is the line-of-sight Galactic acceleration at 3-d position $\bf{x}_i$, $\alosgalmod(\mathbf{r}_i) = \mathbf{a}_{\rm gal}(\mathbf{x_i})\cdot\hat{n}_i$, $\mathbf{a}_{\rm gal}$ is the 3-d galactic acceleration from our previously determined best fit models, and $\hat{n}_i$ is the line of sight unit vector from the solar system to $\mathbf{r}_i$. 

This deviation in the acceleration field from a smooth potential, $\Delta a$, that is experienced by an individual binary pulsar has a high degree of degeneracy between the position of a gravitating mass and the mass itself.  The addition of multiple binaries provides additional constraints and allows us to break this degeneracy to some extent. Hence, we can define the average absolute acceleration deviation for N binaries as:
\begin{equation}
{\Delta \bar{a}} = \frac 1 N \sum_{i=1}^N |\Delta a| (\mathbf{r}_i)|
\end{equation}

An enhanced value for $N>1$ pulsars that are spatially close to each other can indicate the presence of a local enhancement in the mass density.  As we argue in the Appendix, any local enhancement (above some threshold) for which pulsar timing is sensitive to must be due to dark matter sub-halos.

We analyze the largest available dataset of pulsar binaries \cite{Donlonetal2024} for which Galactic accelerations can be cleanly derived and search for sources that display significant excess local acceleration.  For the current set of 27 pulsar binaries for which we have measured Galactic accelerations from the drift rate of the binary period, there are 351 possible pairs of binaries, e.g. $N=2$.  We further require these pairs of binaries are separated by no more than 0.5 kpc as the maximal gravitational sphere of influence for a sub-halo is limited by this (see Appendix).
Thus, we require that the deviation in the acceleration field is both correlated 
(experienced by multiple pulsars) 
and localized.  We impose a localization requirement as excess power can also arise from disequilibrium features on larger (few kpc) scales.  We set the minimal threshold for the deviation in the acceleration field by requiring that it is larger than the maximal local contribution from the gas and stars in the Galaxy.   We show in the Appendix that the maximal contribution to $\Delta a$ from stars and gas in our Galaxy is $\sim 6 \times 10^{-10}~\rm cm/s^{2}$.  
The full pulsar sample that satisfies our selection criteria is given in Table \ref{tab:pulsar_sample}, which notes the binary pairs, the average absolute acceleration deviation, their common center $X,Y,Z$ in Galactocentric coordinates, and their separation. 
As the precision of direct acceleration measurements improves, one can better account for the baryon distribution in the Galaxy to probe even smaller and less massive dark matter sub-halos.  Here, in this first work on measurements of dark matter sub-halos from direct acceleration measurements we impose this threshold acceleration limit to avoid contamination from stars and gas in constraining the properties of the sub-halo.  We find that there is one pair from which we can constrain the properties of the sub-halo (Table \ref{tab:constraints}), and we provide upper limits for the sub-halo masses derived from the other pairs.  

Modern cosmological hydrodynamical simulations performed within the $\Lambda$CDM construct, \cite{Springel:2008by,Hopkins2015,Arora2024}, indicate that dark matter sub-halos of mass $\sim 10^{7}~M_{\odot}$ have scale radii $\sim$ 0.1 kpc.  Such a sub-halo would produce an acceleration of $\sim 10^{-9}~\rm cm/s^{2}$, which is currently detectable from pulsar timing data.  Pulsar timing data have been used to characterize both the smooth component of the Galactic signal ($\approx 3\times 10^{-8}~\rm cm/s^{2}$) \cite{Chakrabartietal2021}, as well as disequilibrium features that impart asymmetries to the line-of-sight Galactic acceleration on several kpc scales \cite{Chakrabarti2020}; these asymmetries are a factor of several smaller than the disequilibrium features in the Galactic acceleration field arising from the smooth component, but are now manifest in the recent pulsar timing data \cite{Donlonetal2024,Donlonetal2025}.  Thus, searching for dark matter sub-structure in direct acceleration measurements - which are expected to produce an acceleration signal of similar magnitude as the disequilibrium features in the Galactic acceleration field but over a more localized region, should now be viable given the current level of precision.  We discuss this in more detail in the Appendix considering the abundance of dark matter sub-halos in cosmological simulations within the Navarro-Frenk-White construct \cite{Bullock_BoylanKolchin2017} and for compact dark matter candidates.

\textit{Results}---We identify five binary pulsar sets that show a local deviation from a smooth potential in their acceleration measurements (above our threshold with a gravitational sphere of influence less than 0.5 kpc).  These pairs are listed in Table \ref{tab:pulsar_sample}, and we note both the magnitude of the deviation as well as the uncertainty.  We note the uncertainty is large for all but one pair of binary pulsars, but will improve in future pulsar timing data releases.  For each such set (each set has two binary pulsars), we performed Monte Carlo Markov Chain (MCMC) calculations similar to earlier work \cite{Chakrabartietal2021,Donlonetal2024} to constrain the properties of the dark matter sub-halo that can produce the level of excess power experienced by the binary pulsar set.  The total acceleration experienced by the binary pulsar at position $\mathbf{r}_i$ is then written as the sum of the contribution from the Galaxy and the sub-halo :
\begin{equation}
a_{\rm los,measured}(\mathbf{r}_i) = \alosgalmod(\mathbf{r}_i) + \alosshmod(\mathbf{r}_i),     
\end{equation}
where $\alosshmod(\mathbf{r}_i)$ is the modelled line of sight acceleration from a subhalo.  This identifies the subhalo acceleration with the deviation from a smooth potential.
Below, we discuss that disequilibrium features imprinted on the Galactic potential would not affect our results (to beyond our current uncertainties).  We do not currently have accelerometers distributed across the physical extent of a sub-halo, so for simplicity we consider the sub-halo to be a compact object, with mass $M_{\rm sub}$ and located at position, $\mathbf{r}_{\rm sub}$.  Thus, we simply write the acceleration produced by the sub-halo to the i-th pulsar as: 
\begin{equation}
a_{\rm los,sub-halo,i}^{\rm mod} = \frac{G M_{\rm sub}(\mathbf{r}_i-\mathbf{r}_{\rm sub})\cdot{\hat{\mathbf{n}}_i}}{|\mathbf{r}_i-\mathbf{r}_{\rm sub}|^{3}}.
\end{equation}

For the Galactic potential, we consider an exponential density profile for the vertical component of the disk, 
\begin{equation} \label{eq:exp_density}
     \rho(z) = \rho_0 \mathrm{e}^{- |z| / h_z}, 
\end{equation} where $h_z$ is the scale height of the disk, and $\rho_0$ is the midplane density of the disk. The vertical component of the acceleration is then given by: 
\begin{equation}
    a_z(z) = -4\pi G \rho_0 h_z \left(1 - e^{-|z|/h_z}\right)\mathrm{sgn}(z),
\end{equation} 
The radial component of the acceleration is calculated assuming a flat rotation curve, where we adopt a local standard of rest velocity from \cite{Quillen2020} ($V_{\rm LSR} = 233.4 \pm 1.4~\rm km/s$) as in earlier work \citep{Chakrabartietal2021}, which employs the
Galactocentric radius of the Sun measured by the GRAVITY collaboration et al. (2018), the proper motion of the radio source associated with $Sgr A^{*}$, and the tangential
component of the solar peculiar motion by \cite{Schonrich2010}.  This value is consistent with the measurement
using trigonometric parallaxes of high-mass star formation
regions from \cite {Reid2019}.  We also include a spiral potential in addition to this axisymmetric component, and we adopt the parameters of the spiral pattern from prior work by \cite{Antoja2011} (their Table 1).  In particular, we adopt two spiral arms, scalelength $R_{\rm \Sigma} = 2.5~\rm kpc$, pitch angle $i=15.5$ degrees, and an amplitude of the spiral pattern $A_{\rm sp} = 1000~\rm (km/s)^{2}\rm kpc^{-1}.$  We have checked that variation of these parameters within the range noted in \cite{Antoja2011} do not affect our results.  We solve for the scale height and mid-plane density of the Galactic potential from the pulsar accelerations as described below.  Additionally, we consider an isothermal disk \citep{BT08}, which has a vertical component of the potential given by: 
\begin{equation}
     \zeta(z) = \sigma^2\ln\left[\cosh\left(\frac{z}{h}\right)\right],
\end{equation}, 
where $\sigma$ is the vertical velocity dispersion of the disk and $h$ is the scale height, which we solve for as described below. We use a flat rotation curve for the radial component of the potential.
We solve for the best-fit parameters for the Galactic potential and the sub-halo simultaneously; these parameters along with the Bayes factors using the Savage-Dickey method \cite{dickey1971} are summarized in Table \ref{tab:Models}.

 We use the python package \texttt{emceee} and adopt flat, uninformative priors for the sub-halo mass and its location, considering a region $<$ 1 kpc in each Cartesian coordinate, and two orders of magnitude in mass.  For the Galactic potential parameters, we similarly adopt flat priors for the exponential and isothermal disk parameters that span two orders of magnitude.
 Of the sets of binary pulsar pairs that show an excess, correlated power larger than our required threshold, only one has $S/N > 3$ in the derived line-of-sight Galactic acceleration, $\rm a_{los,Gal}$.   These are the pairs PSR J1640+2224/PSR J1713+0747.  Each source is part of a binary pulsar pair.  PSR J160+2224 is a binary system composed of a millisecond pulsar and a white dwarf companion on a 175 day orbit \cite{Deng2020}.  PSR J1713+0747 is also a binary system composed of a millisecond pulsar with a white dwarf companion on a 68 day orbit \cite{Splaver2005,Zhuetal2019}.    
 
 The mass of the sub-halo near the common center of the pairs PSR J1640+2224/PSR J1713+0747 is given in Table \ref{tab:constraints}, along with upper limits derived from the other pulsar pairs, modeling the sub-halo as a compact object.  This table is ordered for the binary sets in terms of $S/N$, and we obtain constraints for the pairs PSR J1640+2224/PSR J1713+0747.   Once we have obtained an initial characterization of the sub-halo using the binary pulsar set PSR J1640+2224/PSR J1713+0747, we add binary pulsar sets with $S/N > 0.5$ in this dataset that are near ($<1$ kpc from) the common center of PSR J1640+2224/PSR J1713+0747.    In addition to this pair and those listed in Table 1, this criterion selects for the binary pulsar PSR J1518+4904.  
 We also select the $S/N > 1$ solitary pulsars that are near ($<1~\rm kpc$) the common center of PSR J1640+2224/PSR J1713+0747.  This yields the additional set of solitary pulsars: J0125-2327,J0645+5158,J1012-4235,J1125-6014,J1751-2857,J1910+1256,J2017+0603,J1600-3053.  Fitting to this full set gives $2.45^{+1.07}_{-0.96} \times 10^{7}~M_{\odot}$ for the mass of the sub-halo and we find it is located at Galactocentric coordinates $X = 7.43^{+0.2}_{-0.12}~\rm$ kpc, $Y = 0.38^{+0.11}_{-0.16} ~\rm kpc$, $Z = 0.21^{+0.06}_{-0.11} ~\rm kpc$, where we have modeled the sub-halo as a compact object.  
 
\begin{figure}[h]
\centering
\includegraphics[height=0.4\textheight]{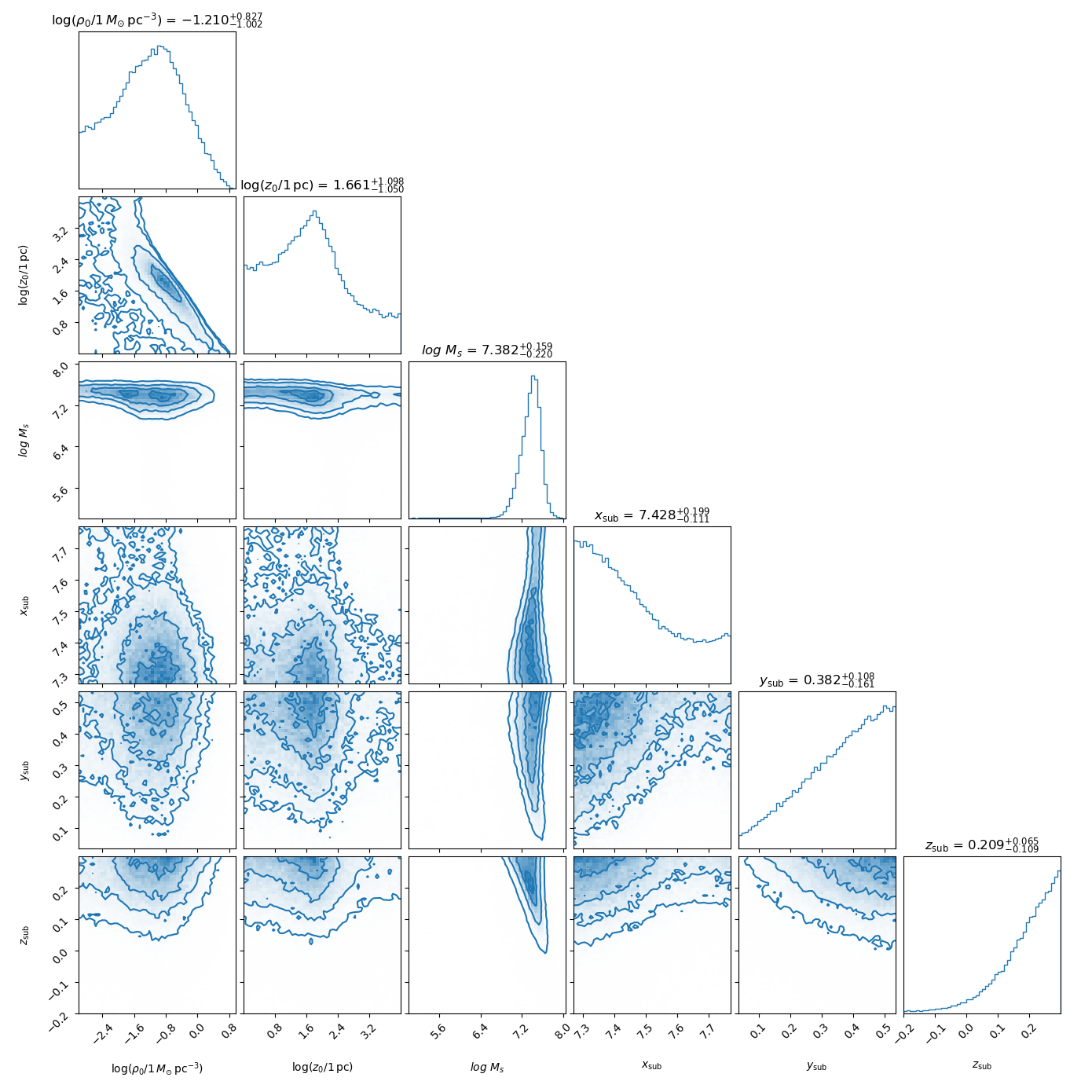}

\caption{{\bf Posterior distribution showing derived properties of a dark matter sub-halo} The panels show the PDFs that display our constraint for the mass of the sub-halo (in solar masses), and its Galactocentric X,Y,Z coordinates using the set of high $S/N$ binary pulsars that are centered near the binary pulsar set PSR J1640+2224, PSR J1713+0747, along with high $S/N$ spin period pulsars.  Here we have modeled the sub-halo as a compact object.  The derived parameters of the mid-plane density and scale height for the exponential disk are also shown.  Here, we show the results for the Galactic potential that includes both the exponential disk and spiral potential (the latter is adopted from \cite{Antoja2011}). 
}
\label{fig:DM1}
\end{figure}

 Figure \ref{fig:DM1} shows the posterior distribution for the mass of the sub-halo, and its location in Galactocentric coordinates for the Galactic potential that is modeled as an exponential disk and a spiral potential.  
 The constraint relies on using more than more binary pulsar, at a minimum the tuple has to include PSR J1713+0747/PSR J1640+2224.  As we discuss in the Appendix, the detection of a $\sim 10^{7}~M_{\odot}$ sub-halo near the Sun is not unexpected in the prevailing $\Lambda$CDM paradigm for a substantial sub-halo mass fraction.  The upper limits from the other binary pulsar sets (given the current precision of the data) indicate that massive ($>10^{8}~M_{\odot}$) sub-halos are disfavored near the Sun.

The Milky Way is now known to have had a highly dynamic history with imprints of dwarf galaxy interactions left on the stars \cite{Widrow2012,Xu2015} and the gas \cite{LevineBlitzHeiles2006,Chakrabarti_Blitz2009,Craigetal2025}.  These disequilibrium features are expected to be manifest on scales of few kpc \cite{Chakrabarti2020} as asymmetries in the acceleration field.  The recent sample of binary pulsars from which Galactic accelerations were derived \cite{Donlonetal2024} indeed shows direct evidence of disequilibrium features at the level of $\sim 10^{-9}~\rm cm/s^{2}$ - over a few kpc scales.  Here, we are analyzing localized $O(1\,{\rm kpc})$ accelerations which would not arise from large-scale disequilibrium features in the Galaxy.


Our fiducial model for the sub-halo consists a compact object and its corresponding acceleration imparted on the pulsars.  The Bayes factors for this model are given in Table \ref{tab:Models} and indicate tentative evidence in favor of a sub-halo.  We have also considered a Navarro-Frenk-White (NFW) density profile for the sub-halo as well.  Here we consider a scale radius of $R_{s}=0.1$ kpc and concentration of 30, motivated by cosmological simulations \cite{Bullock_BoylanKolchin2017}, and fit for $M_{\rm vir}$, using the following mass profile for the sub-halo:
\begin{equation}
M(r) = \frac{M_{\rm vir}}{\rm ln(1+c)-c/(1+c)} \times \left(\rm ln \left(\frac{r + R_{s}}{R_{s}}\right) - \frac{r}{r + R_{s}}\right)
\end{equation}
This case yields a Bayes factor of $\sim$ 30, indicating that it is also favored by the data. In this case, we find a somewhat higher (virial) mass of $6.19^{+1.92}_{-2.03} \times 10^{7}$.  Given that the pulsars that we consider are distributed within a few kpc of the sub-halo and that tidal stripping may be significant for diffuse mass distributions near the Sun \cite{Green2022}, the mass within the scale radius - $0.48^{0.15}_{-0.16} \times 10^{7} M_{\odot}$, may be more relevant.  Unsurprisingly, considering a $r^{-2}$ density profile as indicated in the self-interacting dark matter model \cite{Andoetal2025} for collapsed sub-halos gives a similar Bayes factor as for the compact object acceleration profile.  Given the similar Bayes factors for the compact object and NFW cases, we note that we cannot at the moment determine the density profile of this sub-halo.  

\begin{figure}[h]
\centering

\includegraphics[height=0.25\textheight]{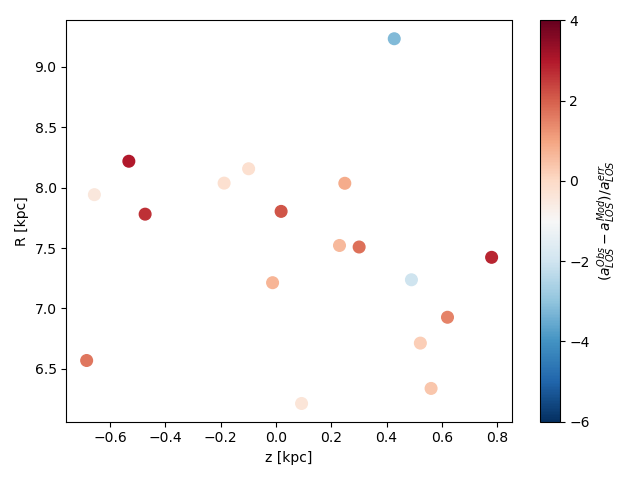}
\includegraphics[height=0.25\textheight]{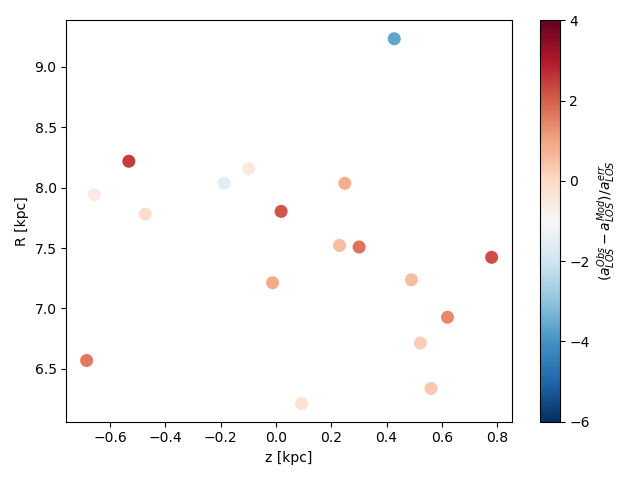}

\caption{The top panel shows the fractional difference in the observed LOS acceleration relative to the model considering an exponential disk in Galactocentric coordinates R,z; the bottom panel shows the same considering an exponential disk and a sub-halo with a NFW profile.  The colorbar shows the fractional difference in the observed and model LOS acceleration, normalized to the error in the observed LOS acceleration; the lighter colors for the sub-halo model suggest that it may be favored by the data.  However, more precise measurements of pulsars (that are clearly above the noise) are needed to confirm this tentative result. 
}
\label{fig:scatter}
\end{figure}

It is important to note that an object with mass $\sim 10^{7}~M_{\odot}/\rm pc^{3}$ with scale radius of 0.1 kpc has a mass density of $\sim 10~M_{\odot}/\rm pc^{3}$, and therefore cannot be in baryonic form; the measured baryonic mass density is $0.084 \pm 0.012~M_{\odot}/\rm pc^{3}$ \cite{McKee2015}.  Figure \ref{fig:scatter} shows the fractional difference between the observed and model LOS accelerations in Galactocentric coordinates, modeling the Galactic accelerations with an exponential disk (top panel), and with an exponential disk and a sub-halo with a NFW profile (bottom panel).  There is some preference in this observable space for the sub-halo model.  We note however that in this initial work, we have not considered a broad range of small-scale baryonic distributions, aside from the spiral arm model; future work should marginalize over a larger range of small-scale baryonic distributions to further assess the sub-halo model.

We are able to constrain a subhalo-like signal in one pulsar pair set. As discussed in the Appendix, under conservative assumptions for the local subhalo fraction ($f_{\rm sub}=0.03$) and subhalo separations from dark matter only simulations, the probability of detecting such a signal in five trials is modest ($\sim 5\%$). However, this estimate assumes randomly distributed search volumes. Our approach is targeted at regions showing correlated excess acceleration deviation, thereby increasing the detection likelihood. Moreover, higher local substructure fractions (e.g., $f_{\rm sub} \geq 0.1$)—well within the range allowed by hydrodynamic simulations—would make such a detection more probable. We note that $f_{\rm sub}$ is still uncertain in simulations at these sub-halo masses; if this detection is robust, simulations can use this measurement as an anchor point for the local substructure fraction near the Sun.
 
 We have examined the \textit{Gaia} photometric data at these coordinates within 1 kpc of each coordinate, and do not find an excess of stars.  Targeted photometric and spectroscopic searches should be carried out to determine if there is an excess of stars in this location that are part of a dwarf galaxy.  Our work here represents the first application and proof of principle of direct acceleration measurements to constrain dark matter sub-structure - future precision measurements - from both pulsar timing, eclipse timing \cite{Chakrabartietal2022,Ebadi2025}, and extreme-precision radial velocity observations \cite{Chakrabarti2020}, should also target this area in the sky to obtain tighter constraints.   As the quantity and precision of acceleration measurements continues to improve, they will provide more definitive constraints on dark matter models.

  \begin{table}[p]
\centering
    \textbf{Pulsar binaries with localized acceleration deviation} \\[0.5ex]
   \begin{tabular}{@{} ccccccccccc @{}} 
      \toprule
      Pulsar pairs & $\Delta a [10^{-9} \rm cm/s^{2}]$ & $\bar{X}, \bar{Y}, \bar{Z}$ [kpc] & $\Delta r$ [kpc] \\
      \hline
      
J1640+2224, J1713+0747 & $7 \pm 4$ & 7.3, 0.57, 0.63
 &  0.36   \\ 


J1455-3330, J1614-2230 & $10 \pm 10$ & 7.51,  -0.21, 0.26 & 0.28  \\ 

J1738+0333, J1741+1351 & $8 \pm 15$ & 6.8, 0.85, 0.56 & 0.30 \\ 

J2145-0750, J2222-0137 & $1 \pm 1$ & 7.9, 0.28,-0.33 & 0.45 \\ 

J2145-0750, J2234+0611 & $4 \pm 27$ & 7.8, 0.5, -0.57 & 0.37 \\


\hline
\end{tabular}
 \caption{Listed are the binary pulsar sets (each set has two binary pairs) used to derive constraints on the sub-halos.  The columns give $\Delta a$, the average absolute acceleration deviation and its uncertainty (the latter is computed in quadrature from the errors for each individual binary pulsar), the average Galactocentric coordinates for the pulsar set, and the $\Delta r$ for the set, which is the radial distance between the pulsar set.  For all but one set of binary pulsar pairs the uncertainties in the localized deviation are currently large.}
 \label{tab:pulsar_sample}
\end{table}


  \begin{table}[p]
\centering
    \textbf{Sub-halo constraints from pulsar accelerations} \\[0.5ex]
   \begin{tabular}{@{} ccccccccccc @{}} 
      \toprule
      Pulsar pairs & M [$10^{7}~M_{\odot}$]  \\
      \hline
      
J1640+2224, J1713+0747 & $2.45^{+1.07}_{-0.96}$     \\  


J1455-3330, J1614-2230 & $< 1.4 $  \\ 

J1738+0333, J1741+1351 & $< 2.5$  \\ 

J2145-0750, J2222-0137 & $<0.56 $  \\ 

J2145-0750, J2234+0611 & $<1.5 $ \\ 


\hline
\end{tabular}
 \caption{Listed are the binary pulsar sets used to derive constraints on the sub-halos.  The derived sub-halo mass is given for the sub-halo centered near J1640+2224/J1713+0747 and upper limits are given for the rest.  The constraints on the sub-halo are derived using not only J1640+2224/J1713+0747 but also the list of binary pulsars and solitary pulsars listed in the text, modeling the sub-halo as a compact object.  Table \ref{tab:Models} includes the mass value when modeling the sub-halo with a NFW profile.}
 \label{tab:constraints}
\end{table}

 \begin{table}[p]
\centering
    \textbf{Models, best-fit values, and Bayes factors} \\[0.5ex]
   \begin{tabular}{@{} ccccccccccc @{}} 
      \toprule
      Model & Parameters & Bayes factor \\
      \hline
      
Exponential disk + spiral & \makecell{$\rm log(\rho_{0}/1~M_{\odot}~ pc^{-3}) = -1.21^{+0.83}_{-1.00}$ \\ $\rm log(z_{0}/1~\rm pc) = 1.66^{+1.09}_{-1.05}$ \\
log $M_{s} = 7.38^{+0.16}_{-0.22}$ \\
$X_{\rm sub} = 7.43^{+0.2}_{-0.12}$  \\ 
$Y_{\rm sub}=0.38^{+0.11}_{-0.16}$  \\ 
$Z_{\rm sub}=0.21^{+0.06}_{-0.11}$} & 38   \\  

\hline 

\makecell{Exponential disk + spiral/ \\
NFW} & \makecell{$\rm log(\rho_{0}/1~M_{\odot}~ pc^{-3}) = -1.37^{+0.79}_{-0.99}$ \\ $\rm log(z_{0}/1~\rm pc) = 1.79^{+0.96}_{-1.22}$ \\
log $M_{s} = 7.79^{+0.17}_{-0.12}$ \\
$X_{\rm sub} = 7.47^{+0.21}_{-0.14}$  \\ 
$Y_{\rm sub}=0.38^{+0.11}_{-0.16}$  \\ 
$Z_{\rm sub}=0.21^{+0.06}_{-0.11}$} & 30   \\  

\hline

Exponential disk/NFW &  \makecell{$\rm log(\rho_{0}/1~M_{\odot}~ pc^{-3}) = -1.16^{+0.89}_{-0.87}$ \\ $\rm log(z_{0}/1~\rm pc) = 1.89^{+1.4}_{-1.05}$ \\
log $M_{s} = 7.88^{+0.12}_{-0.15}$ \\
$X_{\rm sub} = 7.42^{+0.19}_{-0.13}$  \\ 
$Y_{\rm sub}=0.39^{+0.10}_{-0.15}$  \\ 
$Z_{\rm sub}=0.25^{+0.06}_{-0.12}$} & 29   \\ 

\hline

Exponential disk &  \makecell{$\rm log(\rho_{0}/1~M_{\odot}~ pc^{-3}) = -1.17^{+0.84}_{-1.05}$ \\ $\rm log(z_{0}/1~\rm pc) = 1.63^{+1.06}_{-1.04}$ \\
log $M_{s} = 7.45^{+0.15}_{-0.20}$ \\
$X_{\rm sub} = 7.47^{+0.19}_{-0.10}$  \\ 
$Y_{\rm sub}=0.39^{+0.10}_{-0.14}$  \\ 
$Z_{\rm sub}=0.22^{+0.05}_{-0.09}$} & 21   \\ 

\hline 

Isothermal disk & \makecell{$\sigma= 19.7^{+2.12}_{-1.9}~\rm km/s$ \\ $h = 0.62^{+0.97}_{-0.24}~\rm kpc$ \\
log $M_{s} = 7.41^{+0.18}_{-0.23}$ \\
$X_{\rm sub} = 7.42^{+0.19}_{-0.11}$  \\ 
$Y_{\rm sub}=0.39^{+0.10}_{-0.15}$  \\ 
$Z_{\rm sub}=0.22^{+0.05}_{-0.1}$} & 30 \\ 
\hline
\end{tabular}
 \caption{Parameters for Galactic potentials along with derived sub-halo properties (mass of the sub-halo is in solar masses and the location is in Galactocentric coordinates in kpc) are given along with the Bayes factors (using the Savage-Dickey method).  The parameters of the adopted spiral potential are from \cite{Antoja2011}.  The fiducial model for the sub-halo is a compact object, but we also consider a NFW profile.  Unless noted, the sub-halo is modeled as a compact object.}
 \label{tab:Models}
\end{table}

\bigskip
\bigskip 
\begin{acknowledgments}
SC acknowledges support from
NASA EPSCoR CAN AL-80NSSC24M0104, STSCI
GO 17505, and the Margaret Burbidge fellowship at UCSD.
PC acknowledges support from the NASA ATP program through NASA grant NNH17ZDA001N-ATP and National Science Foundation (NSF) via AST-2108269 and AST-2307885. 
This work is partly supported by the U.S.\ Department of Energy grant number de-sc0010107 (SP).  We gratefully acknowledge insightful conversations with Sarah Vigeland, Ethan Nadler, Enrico Ramirez-Ruiz, Manoj Kaplinghat, and especially with Chris McKee that have improved our paper.  We also acknowledge helpful and comprehensive comments from anonymous referees.
\end{acknowledgments}

\clearpage
\bibliography{bibl}

\section*{Appendix}

\textit{Data and samples selection}-- We selected all available binary pulsars from the Australia Telescope National Facility (ATNF) pulsar catalog \cite{Manchester2005} that satisfied the following criteria: the source had a measured $P_b$ and $\dot{P}_b$; had measured parallax $\pi$ and proper motion $\mu$; had measured orbital eccentricity $e$, pulsar mass $m_p$, and companion mass $m_c$.  To enable a clean measurement of Galactic accelerations, we moreover require that the source cannot be a redback, black widow, or transferring mass with its companion, which can change the orbital period of the system.  We also require that the source cannot be in a globular cluster, as the internal accelerations of globular clusters are larger than the Galactic acceleration signal. This yields a total of 27 binary pulsars \cite{Donlonetal2024} that we searched for a deviation from the smooth Galactic potential.  Below we consider possible contamination arising from stars and gas, as well as pulsar planets.

\textit{Accelerations from stars and gas}--To obtain constraints on the differential acceleration arising from stars in the Galaxy near the common center of PSR J1640+2224/PSR J1713+0747, we used \textit{Gaia} sources for which the 3D position can be computed, that is all sources with available sky positions and parallaxes. We selected only stars with positive parallaxes $\varpi$ with relative uncertainty smaller than $20\%$, i.e. the ones satisfying $\varpi/\sigma_{\varpi}>5$. This selection ensures that $1/\varpi$ is a reasonably good estimator of the distance to the stars.  The acceleration signal across the stellar sample is then computed between the binary pulsar pairs PSR J1640+2224/PSR J1713+0747. We use the stellar mass estimates from the Gaia astrophysical parameters table when available, and assume a mass of 1 $M_{\odot}$ otherwise. This mass assumption is rather conservative compared to the typical stellar mass for \textit{Gaia} sources, and is intended to provide an upper limit on the total stellar acceleration from this Gaia sample. We also utilize a Monte Carlo sampling technique to estimate the range of plausible accelerations based on the uncertainties in the \textit{Gaia} data. This resamples across sky position, distance and mass (when available). In each iteration, we recompute the total acceleration.  In addition, we have computed the acceleration that would be experienced with $\sim$ 0.5 kpc of the common center of PSR J1640+2224/PSR J1713+0747 by using the disk component of the \texttt{Galpy} potential \cite{Bovy2015}. This increases the threshold by a factor of $\sim$ two relative to the value determined above using \textit{Gaia} data, but does not change the result significantly.

Using the data from the Leiden-Argentine-Bonn Survey \cite{Kalberla2005}, a recent map \cite{Craigetal2025} of the atomic hydrogen in the Galaxy was constructed that does not rely on kinematic distances, and is more accurate than traditional kinematic maps \cite{LevineBlitzHeiles2006}.  We determine the maximal gas density in the disk from this map \cite{Craigetal2025}, and use this to determine the upper limit on the accelerations that can arise from the gas.  To determine the mass, we integrate the density over a spherical region of size comparable to the distance between the binary pulsars PSR J1640+2224/PSR J1713+0747 ($\sim 0.3$ kpc). This produces a very conservative upper limit for the contribution from the atomic hyrodgen gas disk to the measured acceleration.  We have also examined the CO data near these pulsars, using the dataset compiled by \cite{Rice2016}.  The maximal contribution to the acceleration for the ten closest CO clouds to these pulsars is $\sim 10^{-12}~\rm cm/s^{2}$, i.e., it is much less than the signals that we consider here. The differential acceleration from the atomic hydrogen gas and stars as would be experienced by an object near the common center of J1640+2224/PSR J1713+0747 is at most $6 \times 10^{-10}~\rm cm/s^{2}.$  Here, we analyze the collective effects of multiple pulsars on a local ($\leq$ 0.5 kpc) deviation from a smooth potential, which cannot be produced by planetary systems at these scales.  We thus impose a minimum threshold of $\Delta a = 6 \times 10^{-10}~\rm cm/s^{2}$ (that is a conservative upper limit that could be produced by the baryonic components) in searching for pulsar pairs that may be accelerated by a dark matter sub-halo.

\textit{Abundance of detectable dark matter sub-halos}--The number density of dark matter subhalos as a function of mass is commonly characterized using cosmological dark-matter-only N-body simulations such as Via Lactea and the Aquarius Project.  While these do not account for baryonic effects, which can suppress or disrupt subhalos near the Galactic center, they remain approximately valid for subhalos with masses $\geq 10^{7} M_{\odot}$ at distances $\geq 1$ kpc from the Sun. This is consistent with hydrodynamical simulations \cite{Sawalaetal2017,GarrisonKimmeletal2017,Samueletal2020}, which show survival of massive subhalos even in the presence of a realistic Galactic disk.  In particular, \cite{Sawalaetal2017} find a $\sim$ 50\% reduction in subhalo abundances relative to dark matter only simulations at these radii, which is consistent with \cite{Green2022}.  \cite{WebbBovy2020} report a $\sim$ 25\% reduction in subhalo abundances relative to dark-matter only simulations that is radius-independent.  

We emphasize that there are several factors that currently lead to large uncertainties in subhalo abundances - including artificial disruption in cosmological simulations, differences in subhalo finder algorithms, varying feedback prescriptions in hydrodynamic simulations, and resolution limits.
Most papers reporting strong disruption are based directly on cosmological simulations, which suggests artificial disruption. \cite{Sawalaetal2017} find weaker disruption than other hydrodynamic simulations, potentially due a difference in feedback strength between their work and e.g. FIRE simulations \cite{Hopkins2015}, since the cuspiness of a subhalo strongly affects its susceptibility to disruption.  Idealized simulations \cite{Errani_Navarro2021} find that cuspy subhalos never disrupt. 
\cite{Green2022} and \cite{WebbBovy2020} use hybrid/semi-analytic approaches to integrate subhalo orbits at extremely high resolution.
\cite{Green2022} argue that tidal shocking due to the disk is negligible, and that the disk enhances tidal stripping.  \cite{BensonDu2022} use a semi-analytic model to quantify how much artificial disruption affects inner subhalo counts; in the inner regions, they find an order-of-magnitude difference in counts between current simulations and the "infinite resolution" limit (but the total number of subhalos they predict is low).  Accurately tracking mass loss rates requires extremely high resolution \cite{Mansfield2024}, and current cosmological simulations do not accurately resolve mass loss rates at these distances and subhalo masses. \cite{Mansfield2024} show that subhalo finding algorithms do also matter.  \cite{Wangetal2025} found that the subhalos impacted by the disk do not in fact disrupt (as implied by \cite{GarrisonKimmeletal2017}), but instead are stripped below the resolution limit when using some halo finders. Thus, the accuracy of subhalo halo finders will also affect inferred mass loss rates.  In summary, there are currently a number of factors that lead to \textbf{large} uncertainties in sub-halo fractions in cosmological simulations.  There are suggestions from dynamical modeling of the phase-space spiral found in \textit{Gaia} data that the sub-halo abundance may be higher \cite{Tremaine2023} than found in simulations.

Simulations such as Via Lactea and Aquarius predict a subhalo mass function following approximately
\begin{equation}
\frac{dN}{dM} = A \cdot M^{-\alpha}, \quad \alpha \approx 1.9-2.0
\label{eq:mass_function}
\end{equation}
where $A$ is a normalization constant that depends on the total substructure mass fraction within the region of interest.  The Via Lactea simulation specifically indicates an ``equal mass per decade of mass'' distribution~\cite{Diemand:2007vp}, which corresponds to $M \cdot dN/dM = \text{constant}$, implying $\alpha = 2.0$. We adopt this slope for our analysis. The normalization constant $A$ can be related to the local substructure mass fraction $f_{\rm sub}$ through
\begin{equation}
f_{\rm sub} = \frac{1}{\rho_{\rm total}} \int_{M_{\rm min}}^{M_{\rm max}} M \cdot \frac{dN}{dM} \, dM
\label{eq:mass_fraction}
\end{equation}
The fraction of dark matter contained in substructure with masses $M_{\rm sub} \geq 10^7 M_\odot$ within Milky Way-type halos is constrained to be $f_{\rm sub} \approx 1{-}3\%$ based on high-resolution cosmological $N$-body simulations. The Aquarius project, which simulated six Milky Way-mass halos with up to $\sim 300{,}000$ gravitationally bound subhalos resolved within the virial radius, finds that the cumulative subhalo mass function follows $N(>M_{\rm sub}) \propto M_{\rm sub}^{-0.9}$ \cite{Springel:2008cc}. Similarly, the Via Lactea simulation demonstrates that subhalos are distributed approximately with equal mass per decade of mass, confirming the steep rise in the subhalo mass function toward lower masses \cite{Madau:2008br, Diemand:2008mz}. The relatively low fraction of mass in subhalos above $10^7 M_\odot$ reflects the steep nature of the subhalo mass function, where the majority of substructure mass resides in lower-mass objects \cite{Springel:2008cc, Madau:2008br, Diemand:2008mz}.

For the mass function given in Eq.~\eqref{eq:mass_function} with $\alpha = 2$, this yields
\begin{equation}
f_{\rm sub} = \frac{A}{\rho_{\rm total}} \ln\left(\frac{M_{\rm max}}{M_{\rm min}}\right)
\label{eq:normalization}
\end{equation}
Solving for the normalization constant gives:
\begin{equation}
A = \frac{f_{\rm sub} \cdot \rho_{\rm total}}{\ln(M_{\rm max}/M_{\rm min})}
\label{eq:A_normalization}
\end{equation}
At the Solar position (Galactocentric distance $R_{\odot} \approx 8$~kpc), we adopt the following parameters. We denote the local dark matter density as $\rho_{\rm total}$, with a value of $\approx 0.3$~GeV~cm$^{-3} \approx 5 \times 10^8 M_{\odot}$~kpc$^{-3}$~\cite{Read:2014qva}.  For the minimum subhalo mass, $M_{\rm min}$, we take this to be $10^6 M_{\odot}$ (simulation resolution limit). For the maximum subhalo mass, $M_{\rm max}$, we adopt $10^{10} M_{\odot}$. Finally, we consider a mass range factor: $\ln(M_{\rm max}/M_{\rm min}) = \ln(10^4) \approx 9.2$.  For subhalos with mass $M = 10^8 M_{\odot}$, the number density per unit mass is 
\begin{equation}
\left.\frac{dN}{dM}\right|_{M=10^7 M_{\odot}} = A \cdot (10^7)^{-2} = f_{\rm sub} \times 5.4 \times 10^{-7} \text{ kpc}^{-3} M_{\odot}^{-1}
\label{eq:number_density}
\end{equation}
The number of subhalos within a mass bin $\Delta M$ around $10^7 M_{\odot}$ and within a spherical volume $V$ is
\begin{equation}
N = \left.\frac{dN}{dM}\right|_{M=10^7 M_{\odot}} \times \Delta M \times V = f_{\rm sub} \times 5.4 \times V_{\rm kpc^3}
\label{eq:subhalo_number}
\end{equation}
where we have taken $\Delta M = 10^7 M_{\odot}$ (order-of-magnitude mass bin) and $V_{\rm kpc^3}$ is the volume in units of kpc$^3$.

The mean separation of halos of mass $10^7\ M_\odot$ is thus of order 
\begin{equation}
\Delta r_{\rm subhalo}\simeq (5.4\times f_{\rm sub})^{-1/3}\ {\rm kpc}\simeq 1.8\ {\rm kpc}\  \left(\frac{0.03}{f_{\rm sub}}\right)^{1/3}\left(\frac{M}{10^7\ M_\odot}\right)^{1/3}.
\end{equation}

The pulsar binaries are sensitive to a characteristic acceleration of order $a\sim 10^{-9}\,{\rm cm\,s}^{-2}$.  We can compute the distance to which can sense these effects, $\Delta r_{\rm obs}$:
\begin{equation}\label{eq:distance acceleration}
a = \frac{GM_s}{r^2} \rightarrow \Delta r_{\rm obs} \approx 0.4\left(\frac{M_s}{ 10^7M_{\odot}}\right)^{1/2}\left(\frac{a}{10^{-9}\,{\rm cm\,s}^{-2}}\right)^{-1/2}\,{\rm kpc}
\end{equation}
This motivates our choice searching for pairs and tuplets of pulsars that are separated by $O(1\,{\rm kpc})$.  So for every set of pulsar binaries we are sensitive to a DM subhalo  about $\Delta r_{\rm obs}\simeq 0.4$ kpc away from it.  
The probability to have a subhalo within a sphere of radius 0.4 kpc is, in turn,  
\begin{equation}
P_{\rm subhalo}\approx\left(\frac{\Delta r_{\rm obs}}{\Delta r_{\rm subhalo}}\right)^3\simeq 0.01\left(\frac{f_{\rm sub}}{0.03}\right)\left(\frac{M}{10^7\ M_\odot}\right)^{1/2}.
\end{equation}
The probability of not finding at least one subhalo in five tries is then
\begin{equation}
(1-P_{\rm subhalo})^5\simeq \left[1-0.01\left(\frac{f_{\rm sub}}{0.03}\right)\left(\frac M {10^7\ M_\odot}\right)^{1/2}\right]^5.
\end{equation}
For an enhanced $f_{\rm sub} = 0.1$, there is a 25\% chance of finding at least one subhalo with mass $3\times 10^7\ M_\odot$ in five tries.  This rises to 46\% if $f_{\rm sub} = 0.2$. 
For an enhanced $f_{\rm sub} = 0.1$, there is a 25\% chance of finding at least one subhalo with mass $3\times 10^7\ M_\odot$ in five tries.  This rises to 46\% if $f_{\rm sub} = 0.2$.  We note that these are illustrative estimates rather than firm predictions, and depend sensitively on the (aforementioned) uncertain $f_{\rm sub}$ values.

\textit{Point-like DM candidates}--The number density of macroscopic but point-like (PL) dark matter ``particles'' (e.g. primordial black holes) $M_{\rm PL}$, making up a fraction $f_{\rm PL}$ of the Galactic and cosmological dark matter,  is $n=f_{\rm PL}\rho_{\rm DM}/M_{\rm PL}$; therefore, the typical distance between two such objects is $\approx(n)^{-1/3}=(M_{\rm PL}/(f_{\rm PL}\rho_{\rm DM}))^{1/3}$, or
\begin{equation}
\Delta r_{\rm PL}\simeq 2.7\ {\rm kpc}\ \left(\frac{10^{-3}}{f_{\rm PL}}\right)^{1/3}\left(\frac{M_{\rm PL}}{10^7\ M_\odot}\right)^{1/3},
\end{equation}
implying a probability of 
\begin{equation}
P_{\rm PL}\approx\left(\frac{\Delta r_{\rm obs}}{\Delta r_{\rm PL}}\right)^3\approx 0.01\left(\frac{10^{-3}}{f_{\rm PL}}\right)^{1/3}\left(\frac{M_{\rm PL}}{10^7\ M_\odot}\right)^{1/3}
\end{equation}
i.e. typically one order of magnitude larger than for subhalos. In three tries, it takes a fractional abundance $f_{\rm PL}\simeq 0.03$ of $2\times 10^7\ M_\odot$ point-like masses relative to the local dark matter density. This is in contrast with limits on the abundance of point-like objects in that mass range \cite{Carretal2020,Green2024,Mrozetal2024,Carr2021,Kazunori2014}, unless a large local over-density exists.  For example,  it is possible that this non-baryonic object could be a runaway black hole, as found in other galaxies \cite{VanDokkum2023}.

\end{document}